# THE COLLAPSE OF BELL DETERMINISM

James D. Malley

*Center for Information Technology, NIH, Bethesda MD*

The Bell-Kochen-Specker conditions (*BKS*) for a deterministic noncontextual hidden-variable model are wonderfully simple to state, deal with just one-dimensional projectors on a Hilbert space $H$ and make no reference to a probabilistic phase space or quantum system. They only ask for an assignment of zero or one to every projector such that the assignment respects orthogonal resolutions of the identity. Various no-go results in the literature show that the pair of statements {*BKS* is valid; dim $H \geq 3$} are inconsistent. Here we show, more radically, that the pair actually contradicts the dimensionality of the space itself, by implying that there can exist at most a single one-dimensional projector acting on $H$. Our derivation involves only elementary inner product spaces. It is non-probabilistic, inequality-free, state independent, does not use entanglement, and is simultaneously valid in all dimensions three or greater.

## I. INTRODUCTION

The Bell-Kochen-Specker conditions, *BKS* (also called here Bell determinism), constrain the valuations in noncontextual, deterministic hidden-variable models of quantum events. These easily stated *BKS* conditions for value assignments are elementary geometry in character, and their natural simplicity has lead to a highly evolved, elegant and still developing literature on the search for sets of vectors (projectors) violating the conditions, and thus disallowing one version of a deterministic hidden-variable model; see for example [1]. Here we show that, for $n = \dim H \geq 3,$ the *BKS* conditions imply that the space of projectors collapses to a single one-dimensional projector. This stark outcome thus violates even the dimensionality of the ambient Hilbert space in which the argument takes place. The result therefore provides added information about the lack of relevance of the deterministic *BKS* conditions for quantum outcomes. More precisely, the *BKS* conditions are not merely contradictory in a technical sense but turn out to be deeply at odds with the state space setting itself, and so also with the entire wave function framework of quantum mechanics.

Our derivation involves only elementary inner product spaces. It is nonprobabilistic, inequality-free, state independent, does not use entanglement, and in fact makes no reference to any quantum system. It is simultaneously valid in all dimensions three or more.

The paper is organized as follows. We begin with a discussion of the *BKS* conditions, derive a few vector and affine space lemmas, and then show that only a space containing a single one-dimensional projector is consistent with the *BKS* conditions.

## II. THE *BKS* CONDITIONS

The Bell-Kochen-Specker conditions, *BKS* (also called here Bell determinism), are as follows. We start with Hilbert space $H$ having $n = \dim H \geq 3$, and assume:

(*BKS1*)  there exists a function $v$ on $H$ such that $v(x) = 0$ or $1$, for any vector $x \in H$;

(*BKS2*)   for $\{x_i\}$ any orthogonal basis of $H$, we have $\Sigma v(x_i) = 1$.

We refer to the function $v$ as the *valuation* on $H$.

Now suppose projector $P$ is expressible in two ways as a sum of orthogonal one-dimensional projectors, $P = \Sigma S_i = \Sigma T_i$, for $S_i \perp S_j, T_i \perp T_j, i \neq j$. Then we see that $\Sigma v(S_i) = \Sigma v(T_i)$. As proof, extend $P$ to an orthogonal resolution of the identity using projectors $\{Z_k\}$, so that $I = \Sigma S_i + \Sigma Z_k = \Sigma T_i + \Sigma Z_k$. Using *BKS2* it follows that $\Sigma v(S_i) = \Sigma v(T_i)$. Consequently we obtain an extension of $v$ to an arbitrary projector $P$, using any orthogonal one-dimensional decomposition of $P$, and we can easily show that the valuation $v$ is *orthogonally additive* on projectors: for $x \perp y$, $v(P_x + P_y) =$

$v(P_x) + v(P_y)$. In particular, it follows that if $v(x) = 1$ and $x \perp y$ then $v(y) = 0$. We will routinely use the fact that $v(cx) = v(P_{cx}) = v(P_x) = v(x)$, for any (complex) constant $c \neq 0$.

### III. DEFINITIONS AND LEMMAS

We require a few definitions and simple lemmas for inner product spaces from [2], but our discussion is entirely self-contained. Throughout we assume the *BKS* conditions apply.

*Definition 1.* On inner product space $V$ with $x, y \in V$, define

$$w(x, y) = (2 + \langle x, x \rangle + \langle y, y \rangle)^{-1}[(1 + \langle y, y \rangle)x + (1 + \langle x, x \rangle)y]. \tag{3.1}$$

*Definition 2.* Consider a real subspace $M \subseteq H$, for $\dim M = 3$. Assume $g \in M$, $\langle g, g \rangle = 1$, and define the affine space $S(g) \subseteq M$ by: $S(g) = \{h \in M \mid \langle g, h \rangle = 1\}$.

Observe that $S(g)$ is a two-dimensional hyperplane in $M$, having as "origin" the distinguished point $g$, and therefore has the structure of a vector space. Viewed as an object within $M$ it is an affine space, being a vector translation of a two-dimensional subspace of $H$. If we suppose that $\{g, h_1, h_2\}$ is an orthonormal basis for $M$, then $S(g)$ has a positive definite inner product given by Parseval's relation:

$$\langle \varphi_1, \varphi_2 \rangle_S = \langle \varphi_1, h_1 \rangle \langle \varphi_2, h_1 \rangle + \langle \varphi_1, h_2 \rangle \langle \varphi_2, h_2 \rangle, \text{ for } \varphi_1, \varphi_2 \in S(g). \tag{3.2}$$

When vectors $X, Y \in M$ are viewed as elements of $S(g)$ we write $X \perp_S Y$, $\langle X, Y \rangle_S$, and $w = w_S[X, Y]$. With $X = x + g$, for any $x \in M$, $x \perp g$, we define $v(X) = v(x + g) =$ $v(P_{x+g})$, viewing $x+g \in M$. Note that with $\lambda \in \mathbb{R}$ we have $v(\lambda X) = v(X)$ for $X$ and $\lambda X$ considered as elements of $M$ (since $P_{\lambda X} = P_X$). However, $\lambda X$ considered as an element of vector space $S(g)$ is generally not equal to $\lambda X$ considered as an element of vector space $M$. In particular, it need not be true that $v(\lambda X) = v(X)$, for $X$ and $\lambda X$ considered as elements of $S(g)$. In what follows local context should keep confusion to a minimum.

*Lemmas 1* and *2*, below, are just restatements or restricted versions of inner product space facts given in [2; pp. 121-130]. It is assumed throughout that $\langle g, g \rangle = 1$.

*Lemma 1.* Suppose $v(g) = 1$, and $X, Y \in S(g)$. Then (a) $\langle X, Y \rangle = \langle X, g \rangle \langle Y, g \rangle + \langle X, Y \rangle_S$; (b) $\langle X, Y \rangle_S = -1$ if and only if $\langle X, Y \rangle = 0$; (c) $\langle X, Y \rangle_S = -1$ implies $v(X) + v(Y) = v[w_S(X,Y)]$; (d) $\langle X, Y \rangle_S = 0$ implies $v(X+Y) \leq v(Y)$.

*Proof.* Part (a) is routine, being Parseval's relation for the inner product on $M$, and (b) follows immediately from (a). To prove (c) note that, if $\langle X, Y \rangle_S = -1$, for any $X, Y \in S(g)$, then $\langle X, Y \rangle = 0$, using (b). Next let

$$Z = -\langle Y, w_S(X,Y) \rangle \langle X, w_S(X,Y) \rangle^{-1} X + Y, \tag{3.3}$$

so $Z$ is in the subspace in $M$ generated by $\{X, Y\}$ and check that $Z \perp w_S(X,Y)$. We show $Z \perp g$. For this, note first that $\langle Z, g \rangle = -\langle Y, w_S(X,Y) \rangle \langle X, w_S(X,Y) \rangle^{-1} + 1$. But we also have

$$\begin{aligned}\langle Y, w_S(X,Y) \rangle &= [2 + \langle X, X \rangle_S + \langle Y, Y \rangle_S]^{-1}[1 + \langle X, X \rangle_S] \langle Y, Y \rangle \\ &= [2 + \langle X, X \rangle_S + \langle Y, Y \rangle_S]^{-1} \langle X, X \rangle \langle Y, Y \rangle, \end{aligned} \tag{3.4}$$

and

$$\begin{aligned}\langle X, w_S(X,Y) \rangle &= [2 + \langle X, X \rangle_S + \langle Y, Y \rangle_S]^{-1}[1 + \langle Y, Y \rangle_S] \langle X, X \rangle \\ &= [2 + \langle X, X \rangle_S + \langle Y, Y \rangle_S]^{-1} \langle Y, Y \rangle \langle X, X \rangle. \end{aligned} \tag{3.5}$$

It follows that $Z \perp g$.

Observe next that the space spanned by $\{w_S(X,Y), Z\}$ is identical to the space spanned by $\{X,Y\}$, and recall that $w_S(X,Y) \perp Z$, and $X \perp Y$. Therefore $P_{w_S(X,Y)} + P_Z = P_X + P_Z$. Also, $Z \perp g$ with $v(g) = 1$, so that $v(Z) = 0$. Consequently

$$\begin{aligned} v(X) + v(Y) &= v(P_X + P_Y) = v(P_{w_S(X,Y)} + P_Z) \\ &= v(P_{w_S(X,Y)}) + v(P_Z) = v(P_{w_S(X,Y)}) \\ &= v[w_S(X,Y)], \end{aligned} \qquad (3.6)$$

as required. To prove $(d)$ we need only paraphrase from the first few lines of [2; *Theorem 5.16*]. Thus, the inequality clearly holds if $X = 0$. If $Y = 0$, $X \neq 0$, then using $W = -X/\langle X,X \rangle_S$ we have $\langle X,W \rangle_S = -1$, $w_S(X,W) = 0$, and $v(X+Y) = v(X) = v(0) - v(W) \leq v(0) = v(Y)$. Next, suppose that $X, Y \neq 0$, $\langle X,Y \rangle_S = 0$. Let $W$ be the point on the straight line $\overline{Y(X+Y)}$ satisfying $\langle W, X+Y \rangle_S = -1$. Specifically, let $W = tX + Y$ for $t = -(1 + \langle Y,Y \rangle_S)/\langle X,X \rangle_S$. With a little effort it can be verified that $w_S(W, X+Y) = Y$, and so using $v(W) \geq 0$, we then get $v(X+Y) \leq v(X+Y) + v(W) = v[w_S(W, X+Y)] = v(Y)$, as required. ∎

*Corollary 1.* Suppose $v(g) = 1$, and $X \in S(g)$. If $\lambda \in \mathbb{R}$, $\lambda > 1$, then $v(\lambda X) \leq v(X)$.

*Proof.* Choose any $Y \in S(g)$ such that $Y \perp_S X$, $\langle Y,Y \rangle_S = (\lambda - 1)\langle X,X \rangle_S$. Then check that $(X+Y) \perp_S [(\lambda-1)X - Y]$. Using *Lemma 1(d)* (twice) we find $v(\lambda X) = v[X + Y + (\lambda-1)X - Y] \leq v(X+Y) \leq v(X)$, as required. ∎

*Corollary 2.* Assume $v(g) = 1$. Suppose $y \in H$, $\langle y,y \rangle \neq 0$, $y \perp g$. For $\alpha = 1/\langle y,y \rangle_S$, let

$$X = g + y, \quad X_\alpha = g - \alpha y. \text{ Then } v(X) + v(X_\alpha) = 1.$$

*Proof.* Choose vector $z \in H$, such that $\{g, y, z\}$ is an orthonormal triple and let $M$ be the real space they span. For $S = S(g)$ a calculation shows that $\langle X, X_\alpha \rangle_S = -1$ and $w_S[X, X_\alpha] = g$. Using *Lemma 1(c)* we have $v[w_S(X, X_\alpha)] = v(X) + v(X_\alpha) = v(g) = 1$, as required. ∎

We next present a result that is crucial to what follows, but in full generality does use a tiny bit of calculus, namely L'Hôpital's rule. On the other hand, in obtaining our central inconsistency result below, it will clear that we can specialize the proof to just a small, finite set of vectors, and for each of these we can verify the cosine result (below) by direct calculation. That is, even this fragment of calculus is removable.

*Lemma 2.* Suppose $v(g) = 1$, $X, Y \in S(g)$, $\langle X, X \rangle_S < \langle Y, Y \rangle_S$. Then $v(Y) \leq v(X)$.

*Proof.* We start with the calculus fact that $\lim[\cos(\theta/n)]^n = 1$, as $n \to \infty$, so given

$$\beta_Y = \langle Y, Y \rangle_S, \ \beta_X = \langle X, X \rangle_S, \ \beta_{X,Y} = \langle X, Y \rangle_S, \ \theta = \cos^{-1}[\beta_{X,Y}/\sqrt{\beta_X \beta_Y}] \quad (3.7)$$

there must exist an integer $n$ for which $\alpha = \sqrt{\beta_Y/\beta_X} \, [\cos(\theta/n)]^n \geq 1$. In the real space spanned in $S(g)$ by $\{X, Y\}$ define vectors $Y_0 = \alpha X$, $Y_1, Y_2, \ldots, Y_n = Y$ by means of

$$\text{angle}(Y_i, Y_{i-1}) = \theta/n, \ \sqrt{\langle Y_{i-1}, Y_{i-1} \rangle_S} = \sqrt{\langle Y_i, Y_i \rangle_S} \cos(\theta/n), \ i = 1, 2, \ldots, n. \quad (3.8)$$

We check that $(Y_i - Y_{i-1}) \perp_S Y_{i-1}, 1 \leq i \leq n$, so by *Lemma 1(d)* we see that $v(Y_i) = v[Y_{i-1} + (Y_i - Y_{i-1})] \leq v(Y_{i-1})$. Hence $v(Y) = v(Y_n) \leq v(Y_{n-1}) \leq v(Y_{n-2}) \leq \cdots \leq v(Y_0) = v(\alpha X) \leq v(X)$, where the last inequality follows from *Corollary 1*, and this completes the proof. ∎

Let's consider an example of a cosine chain as above. Let $\{x, y, z\}$ be an orthonormal triple, for which $v(x) = 1$. Let $S = S(x)$, $Y = x + y + \sqrt{2}z$, $X = x - y$. We have $\langle X, Y \rangle_S = -1$, $\langle Y, Y \rangle_S = 3$, $\langle X, X \rangle_S = 1$, $\sqrt{\langle X, X \rangle_S / \langle Y, Y \rangle_S} = 1/\sqrt{3} = 0.57735$, and $\theta = \cos^{-1}[-1/\sqrt{3}] \approx 125.264°$. Now for $n = 4$, we have $[\cos(\theta/4)]^4 = 0.53268$, and for $n = 5$, $[\cos(\theta/5)]^5 = 0.61016$. Hence for these $X, Y$ we require a chain of length $m = n + 2 = 7$, that is $\{Y = Y_5, \ldots, Y_1, Y_0 = \alpha X, X\}$ where $\alpha = \sqrt{3}[\cos(\theta/5)]^5 = \sqrt{3}(0.61016) = 1.05683 \geq 1$. Using this chain in *Lemma 2* we obtain $v(Y) \leq v(X)$. Moreover, if $v(X) = 0$ for $X = x - y$ then there are cosine chains allowing us to conclude that $v(Y) = 0$ for each $Y \in S(x)$ such that $\langle Y, Y \rangle_S > \langle X, X \rangle_S = 1$. This is the key to the next result, which does not appear in [2] but is essential to what follows.

*Lemma 3.* Suppose $v(g) = 1$, for unit vector $g$. Then for any vector $y \in H$ not in the span of $g$ we must have $v(y) = 0$. Equivalently, $P_y \neq P_g$ implies $v(y) = 0$.

*Proof.* If $\langle y, g \rangle = 0$ we are done, since $y \perp g$ and $v(g) = 1$, implies $v(y) = 0$. Consider then the vector $y_1 = y / \langle y, g \rangle$. We have $\langle y_1, g \rangle = 1$. Further, $y_1 \neq cg$ for any complex constant $c$, since $y$ is not in the (complex) span of $g$. To see this note that $y_1 = cg$ implies $1 = \langle y_1, g \rangle = \langle cg, g \rangle = c \langle g, g \rangle = c$, so then $y_1 = y / \langle y, g \rangle = g$, and $y = \langle y, g \rangle g$. But this contradicts $y$ not in the span of $g$.

Now, choose unit vector $z \in H$, $z \perp g$, $z \perp y_1$. Put $\tilde{y} = y_1 - g \neq 0$, so that $\langle \tilde{y}, g \rangle = 0$, and let $M$ be the space spanned over $\mathbb{R}$ by the orthogonal triple $\{g, \tilde{y}, z\}$, with $y_1 = g + \tilde{y} \in S(g)$. Applying *Corollary 2* we get $v(g + z) + v(g - z) = v[w_S(g + z, g - z)] = v(g) = 1$, so that $[v(g + z) = 1 \text{ and } v(g - z) = 0]$, or $[v(g + z) = 0 \text{ and } v(g - z) = 1]$. Observe that $\langle g+z, g+z \rangle_S = \langle g-z, g-z \rangle_S = 1.$ As $v(g+z) = 0$ or $v(g-z) = 0$ we can in either case apply *Lemma* 2 to conclude that $v(Y) = 0$ for any $Y \in S(g)$ such that $\langle Y, Y \rangle_S > 1.$

Introduce two vectors $A^+ = g + \tilde{y}/2 + \sqrt{\lambda}z$, $A^- = g + \tilde{y}/2 - \sqrt{\lambda}z$, where $\lambda = 1 + \langle \tilde{y}, \tilde{y} \rangle / 4.$

Then $\langle A^+, A^+ \rangle_S = \langle A^-, A^- \rangle_S = 1 + \langle \tilde{y}, \tilde{y} \rangle / 2 > 1,$ so that $v(A^+) = v(A^-) = 0,$ and

$\langle A^+, A^- \rangle_S = -1.$ Also $w_S[A^+, A^-] = g + \tilde{y} = y_1.$ Thus $0 = v(A^+) + v(A^-) = v[w_S(A^+, A^-)] = v(g + \tilde{y}) = v(y_1) = v(y),$ as required. ∎

### IV. COLLAPSE OF THE AMBIENT SPACE

We now show that under Bell determinism the ambient space collapses:

*Theorem.* Given the *BKS* conditions suppose $x \in H$ is such that $v(x) = 1.$ Let $h$ be any vector in $H$. Then $h$ is in the space spanned by $x$, and $P_h = P_x.$

*Proof.* Without loss we can take $x$ to be a unit vector. For arbitrary vector $h \in H$ assume that: $h$ is not in the complex span of $x$, and $\langle h, x \rangle \neq 0.$ Put $h_1 = h/\langle h, x \rangle.$ Then $\langle h_1, x \rangle = 1,$ and $h_1 = x + \tilde{y},$ for $\tilde{y} = h_1 - x,$ with $\tilde{y} \perp x, \tilde{y} \neq 0.$ Choose any unit vector $z$ for which $z \perp x, z \perp \tilde{y}.$ Let $M$ be the space spanned over $\mathbb{R}$ by the orthogonal triple $\{x, \tilde{y}, z\},$ and put $S = S(x),$ so that $h_1 = x + \tilde{y} \in S(x).$ By *Lemma* 3, since $h$ is not in the complex span of $x$, we see $v(h) = v(h_1) = v(x + \tilde{y}) = 0.$ Let $\alpha = 1/\langle \tilde{y}, \tilde{y} \rangle_S,$ and observe that $v(x - \alpha\tilde{y}) = 0$ using *Lemma* 3 once more. By *Corollary* 2 we see that $0 = v(x + \tilde{y}) + v(x - \alpha\tilde{y}) = v[w_S(x + \tilde{y}, x - \alpha\tilde{y})] = v(x) = 1.$ This is a contradiction, so that *either h* is in the span of $x$, or $\langle h, x \rangle = 0,$ for any vector $h \in H.$ Suppose $\langle h, x \rangle = 0$ and consider the vector $t = h + x.$ Then $\langle t, x \rangle = 1,$ and $t$ is not in the span of $x$. Replacing $h$ by $t$ and arguing as above leads to another contradiction, so the only allowed outcome is that *h* must be in the span of *x*.  ∎

## V. Discussion

We first observe that to obtain a simple inconsistency of the *BKS* conditions, a careful accounting shows that only a finite number of vectors need be invoked to apply any of the lemmas or corollaries. Thus, for a given, specific vector *h* only a finite set of vectors is required to obtain the result $h = cx$, for that vector *x* such that $v(x) = 1$. In this sense our main result is finitely constructive.

Next, it is interesting to witness the residual mix of randomness and determinism that remains under the collapse. The experimenter can select any orthonormal basis at random, and then effectively Nature makes the assignment $v(x) = 1$ for some vector *x* in this basis. Thereafter, all other vectors not in the (complex) span of *x* must be assigned the value zero (by *Lemma* 3). But even this narrow outcome can be sharply refined since, by the *Theorem* above, the collection of all vectors not in the span of *x* is exactly the empty set. Equivalently, following the experimenter-Nature sequence above, there is, under Bell determinism, just one allowed projector: $P_x$.

Summarizing, we have found that the seemingly innocent deterministic hidden-variable model given by the *BKS* conditions has stark consequences: we start with $n = \dim H \geq 3$ but find, using elementary vector space arguments, that there can be at most only a single projector acting on *H*. Given that Bell determinism collides head on with the basic framework for doing quantum mechanics, as we have brought out here, it is difficult to see any useful outcome for physics in further exploration of this form of determinism. Of course this does not affect the interest for physics in exploring variants of the *BKS* conditions, that is, in exploring different kinds of deterministic models, or indeed, exploration of contextual models or suitable hybrids.

*Acknowledgment:* We would like to warmly thank Arthur Fine for his technical insight and encouragement on this project.

## References

[1] A Cabello, JM Estebaranz, and G Garcia-Alcaine. *Phys. Lett. A* 339 (2005) 425.

[2] SG Gudder. **Stochastic Methods in Quantum Mechanics**. North Holland, New York. (1979).